\pdfoutput=1
\documentclass[reprint,amsmath,amssym,aps,prx]{revtex4-2}

\usepackage{graphicx}
\usepackage{dcolumn}
\usepackage{bm}

\begin{document}

\preprint{APS/123-QED}

\title{Neurodegenerative damage reduces firing coherence \\in a continuous attractor model of grid cells }

\author{Yuduo Zhi}
 \email{ydzhi@ucdavis.edu}
\author{Daniel Cox}%
 \email{cox@physics.ucdavis.edu}
\affiliation{%
 Physics Department, University of California, Davis
}%

\date{\today}

\begin{abstract}
Grid cells in the dorsolateral band of the medial entorhinal cortex(dMEC) display strikingly regular periodic firing patterns on a lattice of positions in 2-D space. This helps animals to encode relative spatial location without reference to external cues. The dMEC is damaged in the early stages of Alzheimer's Disease, which affects navigation ability of a disease victim, reducing the synaptic density of neurons in the network.  Within an established 2-dimensional continuous attractor neural network model of grid cell activity, we introduce neural sheet damage parameterized by radius and by the strength of the synaptic output for neurons in the damaged region. The mean proportionality of the grid field flow rate in the dMEC to the velocity of the model animal is maintained, but there is a broadened distribution of flow rates in the damaged case. This flow rate-to-velocity proportionality is essential to establish coherent grid firing fields for individual grid cells for a roaming animal.  When we examine the coherence of the grid cell firing field by studying Bragg Peaks of the Fourier transformed lattice firing field intensity in both damaged and undamaged regions, we find that for a wide range of damage radius and reduced synaptic strength that for undamaged model grid cells there is an incoherent firing field structure with only a single central peak.  In the radius-damage plane this is adjacent to narrow bands of striped lattices (two additional Bragg peaks), which abut an orthorhombic pattern (four additional Bragg peaks), that abut the undamaged hexagonal region (six additional Bragg peaks).  Within the damaged region, grid cells show no Bragg peaks outside the central one which shows reduced intensity with increasing damage, and outside the damaged region the central Bragg peak strength is largely unaffected.   There is a re-entrant region of normal grid firing fields for very large damage area.  We anticipate that the modified grid cell behavior can be observed in non-invasive fMRI imaging of the dMEC. 
\end{abstract}

\maketitle


\section{Introduction}

\begin{figure*}[htb]
\includegraphics[scale=0.15]{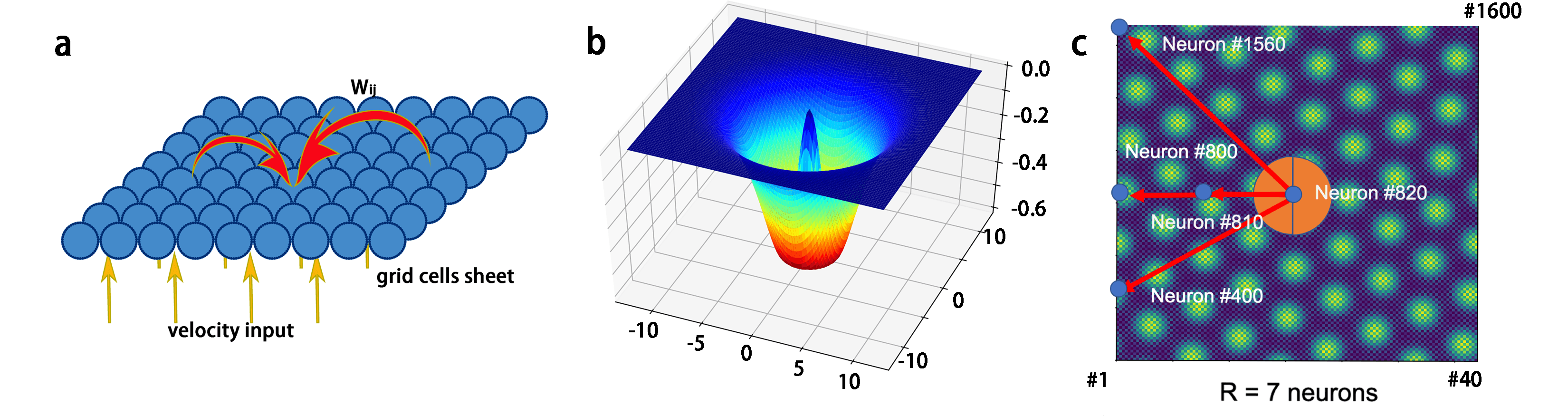}
\caption{\label{fig1} {\bf 2D neuron sheet and damage model.} (a) Blue spheres are neurons in the grid cell model, and red arrows indicate synaptic connections among neurons, with a weight $W_{ij}$ coupling neuron $i$ and neuron $j$. The yellow arrows below are the velocity signal from other cells. The instantaneous velocity input is uniform for the grid cell sheet, but each cell has a different preferred direction. (b) The ``Mexican Hat'' weight matrix $W_{ij}$ is the difference between two Gaussians. It is negative everywhere expect at the center (zero). (c) Central damage model. The heat-map indicates the $40 \times 40$ grid cell layer firing peaks, with the orange circle indicating the damaged region (radius $R$ = 7 neurons in this case). All neurons are numbered from 1 to 1600, and the neuron at the damage center is \#820.}
\end{figure*}

There is considerable interest in understanding how the brain encodes location and guides animal navigation.  Different neural networks within the brain with various functions help to build animals’ navigation system. For instance, place cells in the hippocampus are confirmed to fire strongly at special locations such as reward sites or for the position of external landmarks~\cite{o1978hippocampus,muir2001instability}. Head-direction cells found in many brain areas (e.g., the dorsal presubiculum)~\cite{chen1994head} fire in 1:1 correspondence with the animal’s directional heading with respect to the environment in the horizontal plane~\cite{taube1990head,seelig2015neural}.  The stunning discovery of grid cells in 2005 showed that these neurons in the dorsocaudal medial entorhinal cortex (dMEC) provide an internal coordinate system encoding absolute position for a given enclosure (longitude and latitude) largely independent of external environmental cues~\cite{hafting2005microstructure,mcnaughton2006path}. Each grid cell in a given layer of the dMEC shows enhanced activity (firing) on a periodic hexagonal lattice of points in 2-D space, with the spacing varying with layer depth. Additionally, there is now over a decade of direct evidence that functional Magnetic Resonance Imaging (fMRI) can detect the six-fold symmetry of the grid cell firing pattern noninvasively in healthy brains\cite{Doeller2010,Behrens2016,Behrens2017,Doeller2018}.

Grid cells represent a fascinating example of emergent pattern formation in a nonlinear dynamical system (the coupled neurons of the dMEC).   As such, they are of intrinsic interest within the physics of dynamical systems, and amenable to study and characterization by techniques typically reserved for solid state matter, such as diffraction analysis by Fourier transforms (Bragg peaks).  How such patterns hold up under perturbation is also of intrinsic interest.  

In the case of the dMEC, strong perturbation arises from Alzheimer’s disease (AD), which affects the hippocampus (place cells) and entorhinal cortex (grid cells) and thus can disrupt spatial navigation. Several competing hypotheses exist to explain the cause of the disease. The “tau hypothesis”, proposes that abnormalities associated with tau protein aggregates initiate the disease cascade~\cite{mudher2002alzheimer}. In this model, hyperphosphorylated tau does not hold microtubules together and begins to pair with other threads of tau to form neurofibrillary tangles inside nerve cell bodies~\cite{goedert1991tau}. Because the tau protein is what stabilizes the microtubule bundles in neuronal axons on which neurotransmitters and other cargoes relevant for normal synaptic function are transported,  this may result first in malfunctions in biochemical communication at the synapses between neurons and later in the death of the cells~\cite{chun2007role}. Furthermore, the tau tangles that disrupt the axon and synapses may propagate within the brain from location to location, in a manner similar to the prion protein aggregates of mad cow disease~\cite{wilesmith1988bovine,casalone2018atypical,darwin1873origin}. In particular, the synaptic output will be degraded by axonal  microtubule disruption from tau tangles. The other leading candidates for initiation of Alzheimer's disease, the ``amyloid cascade'' hypothesis~\cite{StrooperKarran2016}, or the related ``amyloid oligomer'' hypothesis~\cite{ClineBicca2018}, lead to eventual tau aggregation as well as an end stage.  

There is substantial direct evidence for AD related damage to the MEC.  Direct post-mortem examination shows significant atrophy of the EC in the brains of AD victims vs. control~\cite{VanHoesen1991}.  fMRI imaging of the MEC region for patients predisposed to early onset AD shows a disruption in the six-fold symmetric firing pattern with respect to the control group when performing virtual navigation tasks, despite no apparent cognitive deficits in the AD disposed group~\cite{Kunz2015}. Similar works on aging adults suggest an impact of AD on the grid cell function and ability to navigate~\cite{Stangl2018}.  Amyloid beta oligomers induce tau tangles in cell culture experiments that degrade microtubules and synaptic quality~\cite{Zempel2013}.  Overexpression of human tau protein with subsequent aggregation in rats leads to degradation of synaptic plasticity in the MEC and degrades cognitive performance~\cite{ChenLiu2018}, and induced expression of mutant human tau in mice leads to grid cell dysfunction~\cite{FuRodriguez2017}.  Finally, there is direct evidence of AD induced synaptic degradation in the neurons projecting from the MEC to the CA1 layer of the hippocampus~\cite{YangYao2018}.  

In this paper, we analyze the impact to animal navigation caused by grid cell damage via neurodegeneration, in which we assume that the synaptic strength of output connection, $\alpha$, with $0\le\alpha\le 1$, for neurons in a circular disk of radius $R$ are degraded, presumably due to the tau tangle influence on the synaptic cargo transport and degradation of microtubules.  Within the continuous attractor model for grid cells introduced by Burak and Fiete~\cite{burak2009accurate}, we find that the damage does not influence the key proportionality between firing flow and animal velocity necessary to achieve path integration and a hexagonal lattice firing pattern, but rather disrupts the coherent pattern of grid cell firing.  That coherence in translation and orientation is readily captured by studying the Fourier transform of the real space firing pattern for each grid cell, i.e., the Bragg peaks corresponding to the grid cell receptive fields in real space. This  yields a phase diagram of coherence in the $\alpha-(1/R)$ plane, with a large region possessing no coherent firing, abutted by a stripe region (two non-trivial Bragg peaks), which is next to an orthorhombic symmetry phase (four non-trivial Bragg peaks), and then the normal coherent grid cell pattern (six hexagonally arrayed non-trivial Bragg peaks).  For cells within the damaged region, the central Bragg peak intensity diminishes uniformly as $\alpha$ is decreased, while it is little affected outside the damaged disk.  There is a mathematically required ``re-entrance'' to the normal phase as $R\to\infty$. We have studied this behavior on single sheets with small numbers of neurons (1600), even though the real medial entorhinal cortex is likely to possess millions of neurons\cite{West1998} and multiple layers\cite{mcnaughton2006path}.  This is of course for convenience of modeling, and we expect our simulations to scale proportionally to the full MEC.    

We note that our phases should be observable in fMRI studies of the MEC.  The previously referenced fMRI studies show clear evidence of the hexagonal symmetry firing in the MEC and disruption of that firing by damage.  This is not surprising in that the EC volume overall is usually $>~1000~mm^3$ or about 1000 fMRI voxels\cite{Price2010}.  We anticipate that this damage should disrupt the coding of position in grid cell/place cell networks\cite{Sreenivasan2011,Sanzeni2016}, but in this paper we restrict our discussion to the damage dependent symmetry modifications and coherence reduction of grid cells which can lead to non-invasive early identification of Alzheimer's or other neurodegeneration that affects the MEC.  

\begin{figure*}[htb]
\includegraphics[scale=0.15]{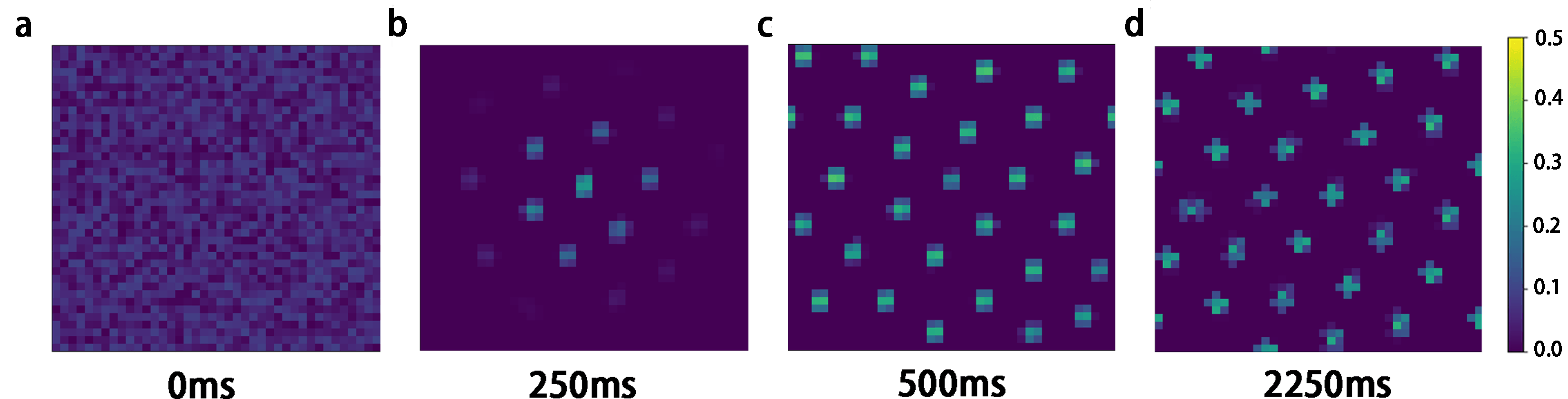}
\caption{\label{fig2} {\bf Temporal emergence of Firing Patterns for undamaged grid cell layer.} (a) Initial state of neuron sheet includes a random noise signal ranging from -0.1 to 0.1. (b) The aperiodic boundary condition shapes the neuronal signal pattern to generate grids in the first 250ms, beginning with an intense central peak with weaker surrounding peaks. (c) Change of the boundary condition from aperiodic to periodic expands the grids evenly. (d) Non-zero velocity inputs ($0.8m/s$ in three directions) heal the defects of previous grids and generate a hexagonal lattice of grid firing peaks. Figures (a)~(d) are heat-maps with the same colorbar.}
\end{figure*}

\begin{table*}[htb]
\caption{Coefficients Table}
\centering
\begin{tabular}{ccccccccc}
\hline
$a$ & $\lambda$(neurons) & $\beta$(1/neuron$^2$) & $\gamma$(1/neuron$^2$)  & $l$(neurons) & $\tau$(ms) & $dt$(ms) & $\eta_0$ & $\mu$ \\
\hline
$1$ & $8$  & $ 3/\lambda^2 $ &$6.711\times\beta$ & 1 & 10 & 0.5 & $0.10315$ &$0.875$ \\
\hline
\end{tabular}
\label{tab1}
\text{ $\lambda$ is the target periodic wavelength of the  triangular lattice, $\eta_0$ in Eq.~\ref{eq4},  $\mu$ in Eq.~\ref{eq7} }
\end{table*}

\section{Methods}
\subsection{Emergent Grid-like Firing Pattern in a Continuous Attractor Model of a Neuronal Sheet.}

In continuous attractor models, each neuron receives inhibitory input from a surrounding ring of local neurons, and the entire network receives broad-field feed-forward excitation containing location and direction data from elsewhere in the brain.  The model, upon integration in a static limit, will tend towards a stable fixed point of the coupled equations, i.e., an attractor in the full phase space of the coordinates. When the model animal is moving, given sufficiently rapid response of the neurons in the model, the stable firing pattern can flow in response to the motion and this is the origin of the observed grid cell pattern from this picture.  We have based our work upon the continuous attractor model of Burak and Fiete\cite{burak2009accurate}, which allows for modest recurrent excitatory synapses between neurons locally surrounded by broadly recurrent inhibitory synapses around a given cell. The model is attractive to use for our purposes here since: (i) it does develop a grid cell like firing pattern in the model sheet, and (ii) with the addition of velocity sensitive response mimicking the input from other parts of the brain (such as from head direction cells) it develops a pattern flow that leads to accurate path integration and a real-space hexagonal firing pattern.  However, a purely inhibitory ring is sufficient to obtain grid cells and this is justified by experimental evidence from studies on rats\cite{Couey2013}. The mixture of excitatory and inhibitory inputs is an attempt to capture in one model a bipartite entorhinal cortex layer containing both excitatory pyramidal cells and inhibitory interneuron cells.

Consider a network of neurons arranged with uniform density on a cortical sheet(Fig.~\ref{fig1}a), and with a connection strength that decreases with distance. If the connections of inhibitory cells extend over a wider range than the connections of excitatory cells, it is possible for an emergent symmetry breaking of the firing pattern with a population response consisting of a regular pattern of discrete regions of neural activity can be created, arranged on the vertices of a periodic structure. As analyzed in the supporting information S4, the most stable steady state structure in undamaged conditions has hexagonal lattice symmetry~\cite{o1978hippocampus,muir2001instability}. 

 The blue spheres in Fig.~\ref{fig1}a represent grid cells in a 2D neuron sheet, corresponding to one of the grid cell layers of the dMEC. For fast simulation, we use a 40$\times$40 neuron sheet, and the coordinates on that 2D plane can be described by a neuron position vector $\vec x_i$. 
 
The dynamics of grid cell activity in this model are described by the coupled differential equations~\cite{dayan2003theoretical,burak2009accurate}
\begin{equation}
\tau\frac{ds_i}{dt}=-s_i+f( \sum W_{ij}\cdot s_i +B_i )~~.
\label{eq1}
\end{equation}
$s_i$ is the $i$-th neuron's firing rate. $\tau$ is the time constant, chosen here to be 10ms, and $B_i$ is the feed-forward input to neuron $i$ (Fig.~\ref{fig1}a,b), which is explained in detail later. The neural transfer function assumed here, per Burak and Fiete~\cite{burak2009accurate}, is a simple rectification non-linearity:  $f(x) = x$ for $x>0$ and 0 otherwise. $W_{ij}$ is the synaptic weight from neuron $j$ to neuron $i$, which has the character that inhibition by neurons operates at longer range than activating ones. Following Burak and Fiete~\cite{burak2009accurate}, the weight matrix function is written as the difference of two Gaussian curves with different variances which has a “Mexican Hat” shape in position space(Fig.~\ref{fig1}b):
\begin{equation}
W_{ij}=W_0(\vec x_i- \vec x_j-l \hat e_{\theta _{j}} )
\label{eq2}
\end{equation}
with 
\begin{equation}
W_{0}(x)=ae^{-\gamma \left| x\right|^2}-e^{-\beta \left|x\right|^2}~~.
\label{eq3}
\end{equation}

In Eq.~\ref{eq2}, we shift the neuron separation by a new term $l \hat e_{\theta _{j}}$. The neuron preferred direction is  $\hat e_{\theta _{j}}$ and we will always choose a non-zero $l$. This shifted location term plays an important role in driving a statistical flow of the grid cell firing pattern, which is explained in Supporting Information S2 in detail.  
The weight matrix function $W_{0}(x)$ in Eq.~\ref{eq3} is the difference of two Gaussians: 1) $a$ is chosen to be 1 to make the net response inhibitory, so the value at the center in Fig.~\ref{fig1}b is zero. A small $a$ is enough to create grid-like firing pattern while $a >1$ would not affect the result qualitatively. There is a relative excitatory response at small separation compared to the maximum inhibition. 2) $\beta$ is used to determine the width of the inhibitory response in the surrounding neuron region, and $\gamma$ sets the shorter distance for the excitatory response in the surrounding neuron region. 3) For the smaller 40$\times$40 system we take $\gamma=6.711\times\beta$ to make the maximum inhibition big enough to generate a grid-like firing pattern for a small dimension lattice. The simulation parameters are listed in Table~\ref{tab1}, in which we have introduced $\lambda$, the target periodic wavelength of the formed triangular lattice~\cite{burak2009accurate}, and we use it to choose $\beta$ and $\gamma$. An approximate relationship is $\lambda\approx\sqrt{3/\beta}$. Using $\lambda$ made it easier to control the firing pattern lattice spacing and it is explained in Supporting Information section S4.  

We employ $40\times 40$ neuron sheets to simulate one layer of of the dMEC, but as noted in the introduction we expect a much larger number in the layers of the dMEC, potentially up to a million per layer\cite{West1998}.  The smaller neuronal lattice is chosen purely for computational convenience, and proportional damage compared to the corresponding damage in the full dMEC, i.e., for $R=10$ so the area is about 300 in model neuronal spacing units, and this would correspond to about 18\% of the neurons in a layer being damaged.  

We start each simulation with small random noise within the range from -0.1 to 0.1 (arbitrary units, but referenced to the static background input of 1), (Fig.~\ref{fig2}a), then apply the aperiodic boundary condition for a 250$ms$ stimulation process. For shorter times, we see randomly separated firing peaks emerge on the inactivated black background (Fig.~\ref{fig2}b), and the central activity in these peaks is higher than the surroundings. Then as we run another 250$ms$ simulation under periodic boundary conditions, we develop a complete periodic lattice of firing peaks (Fig.~\ref{fig2}c). All the simulations above are done with zero velocity input, and that explains why the firing regions are small radius “peaks” instead of the peak clusters (Highlight parts in Fig.~\ref{fig2}d). Then we let the whole neuron sheet complete building triangular lattice using a annealing process: we apply a nonzero velocity input (0.8 $m/s$) in three directions $(0,\frac{\pi}{5}, \frac{\pi}{2}-\frac{\pi}{5})$, and complete a 500$ms$ simulation for each of directions. The annealing process removes the defects and generates a complete triangular lattice (Fig.~\ref{fig2}d)

\begin{figure*}
\includegraphics[scale=0.15]{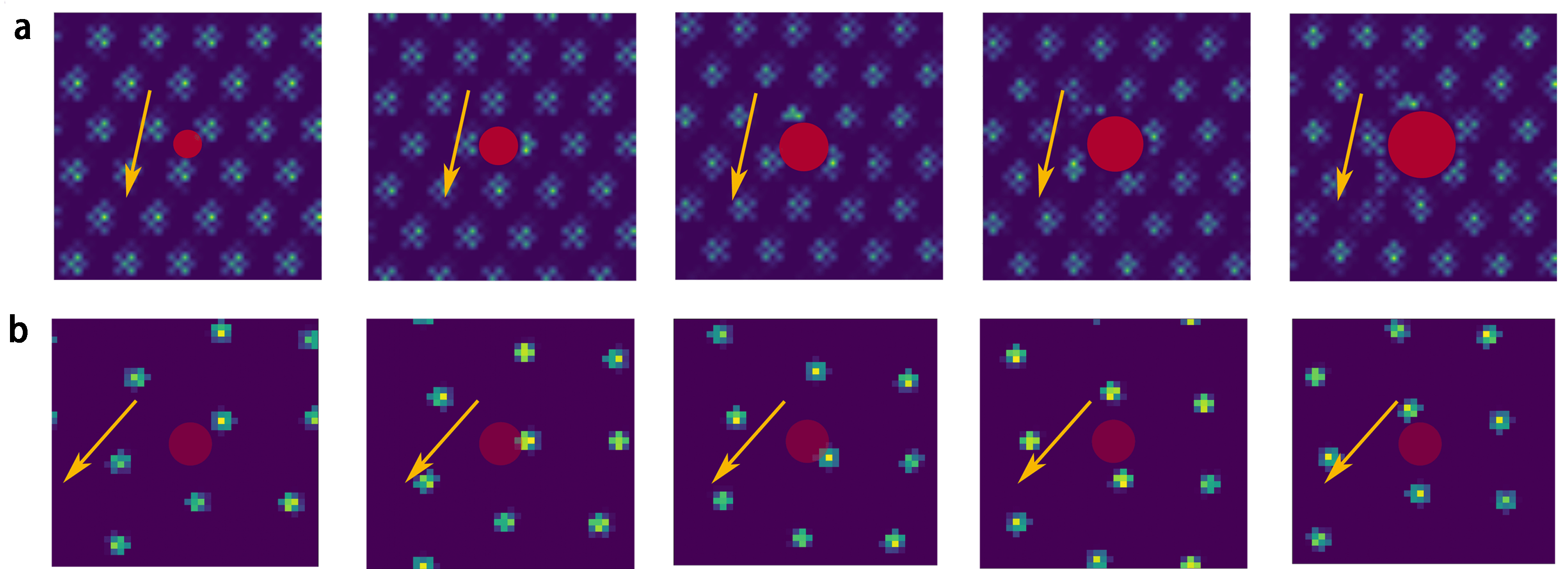} 
\caption{\label{fig3} {\bf Neuron Sheet Firing Pattern with Model Damage.} (a) Dead neurons ($\alpha$=0) in the red damaged regions, with successive damage region radii of 2,4,6,8 neurons. Yellow arrows indicate the flow direction in the opposite direction of the animal velocity, with grid cell firing peaks bypassing the central damage even as it grows. (b) Weakened neuronal firing in the central damage region (red), with $R = 4$ neurons and $\alpha = 0.6 $,The whole grid-like firing pattern is moving along the yellow direction, and neurons fire (more weakly) in the damaged region.}
\end{figure*}

\subsection{Grid-like Firing Pattern Flow}
The shifted location term $l\hat e_{\theta _{j}}$ is associated with the neuron’s preferred direction $\hat e_{\theta_{j}}$, and these orientation sensitive firings drive the grid pattern from stationary to flowing.  This is a way to mimic the input in the model from the head direction cells. In the head direction system, cells fire selectively with respect to the rat’s head orientation as a result of neural integration of head angular velocity signals derived from the vestibular system. In the grid cell system, each neuron receives input from one head-direction cell tuned to its preferred direction, and the neuron’s outgoing center-surround connectivity profile is not centered on itself,  but is shifted by a few neurons along its preferred direction, which is shown in the weight function above as the shifted location vectors.

In our grid-like sheet, we tiled the neuron uniformly in this way: each neuron $i$ has a preferred direction $(W, E, S, N)$, indicted by $\hat e_{\theta _{i}}$, and each 2$\times$2 neuron block contains all four preferred directions. and then we can define the feed-forward input to neuron $i$ is:

\begin{equation}
B_i(x)=A_i(x)(1+\eta_0\hat e_{\theta _{i}}\vec{v}) 
\label{eq4}
\end{equation}

where $\vec{v}$ is the velocity of the rat, in units of $m/s$. $\eta_0$ is the coefficient that characterizes the effects of velocity inputs to the driven pattern flow (in Table~\ref{tab1}).  $A_i (x)$ is called the {\it envelope function} which helps to modulate the strength of the input to the neurons. We assume periodic boundary conditions for firing on the sheet, and in this case
\begin{equation}
A_i(x)=1
\label{eq5}
\end{equation}
\noindent
We have used neuron sheet of a size 40$\times$40 (1600 neurons) to speed up numerical simulations.
And $A_i(x)$ of aperiodic boundary conditions is given in Supporting Information S5.

If we have a non-zero value for the shifted location vector and, then the feed-forward input $\textbf{B}$ will drive a flow of the formed pattern. $\eta_{0}$ determined the gain of the velocity response of the network, and the term $\eta_{0}\hat e_{\theta _{j}}\vec{v}\ll 1 $ stabilizes the flowing lattice. In Supporting Information S2 we explain how the feed-forward input drives flow and S3 shows the influence of $\eta_{0}$.

\subsection{Central Damage Model}
As discussed in the introduction, Alzheimer’s disease affects the hippocampus (place cells) and the entorhinal cortex (grid cells) early and thus disrupts navigation. It may proceed by diffusion of “tau tangles” from cell to cell which will disrupt synaptic function. We focus on one type of damage to the dMEC that can arise from neurodegenerative diseases and affect grid cell performance: diffusing damage that can arise from propagation of neurofibrillary tau tangles similar to the prion diseases~\cite{Hall2012}. Based on this model, we model neuronal functional loss as a weakening of the output synaptic strength, which would follow from tau tangle driven disruption and damage to the axonal microtubule bundles. We do not explicitly model tau tangles in this paper. As we dial the output strength to zero, we effectively ``kill'' the neuron in the model. In Fig.~\ref{fig1}c, we show a central diffusion damage on top of the grid-like firing pattern. The $40\times 40$ healthy neuron sheets has its own triangular grids of firing pattern, and then the neurons within the orange region are set to be damaged, after which we observe neuron signals in different locations( \#820 is within the damage region and \#400, 800, 810, 1560 are healthy neurons) and with different sizes of central damage (damage region radius $R$ = 7 neurons in Fig.~\ref{fig1}c).

As the first example (Fig.~\ref{fig3}a), we kill a central neuron and allow the neuronal damage to propagate outward to model the prion like spread alluded to in the above paragraph. The time step for numerical integration in our simulation is 0.5 $ms$ and we find that 250 $ms$ total integration time (500 time steps) is sufficient for the surrounding neurons to develop a stable firing pattern. These timescales are obviously significantly accelerated from the AD scale so we can see the effects in a reasonable simulation time frame.  The resulting growth speed of the damage cluster radius is 1 neuron/step. We take screenshots every 50 $ms$  to record the damage as we change from a circular damage region with a radius of $R$ = 2 neurons to $R$ = 7 neurons to see the flow on damaged neuron sheets. 
\begin{figure}
\includegraphics[scale=0.123]{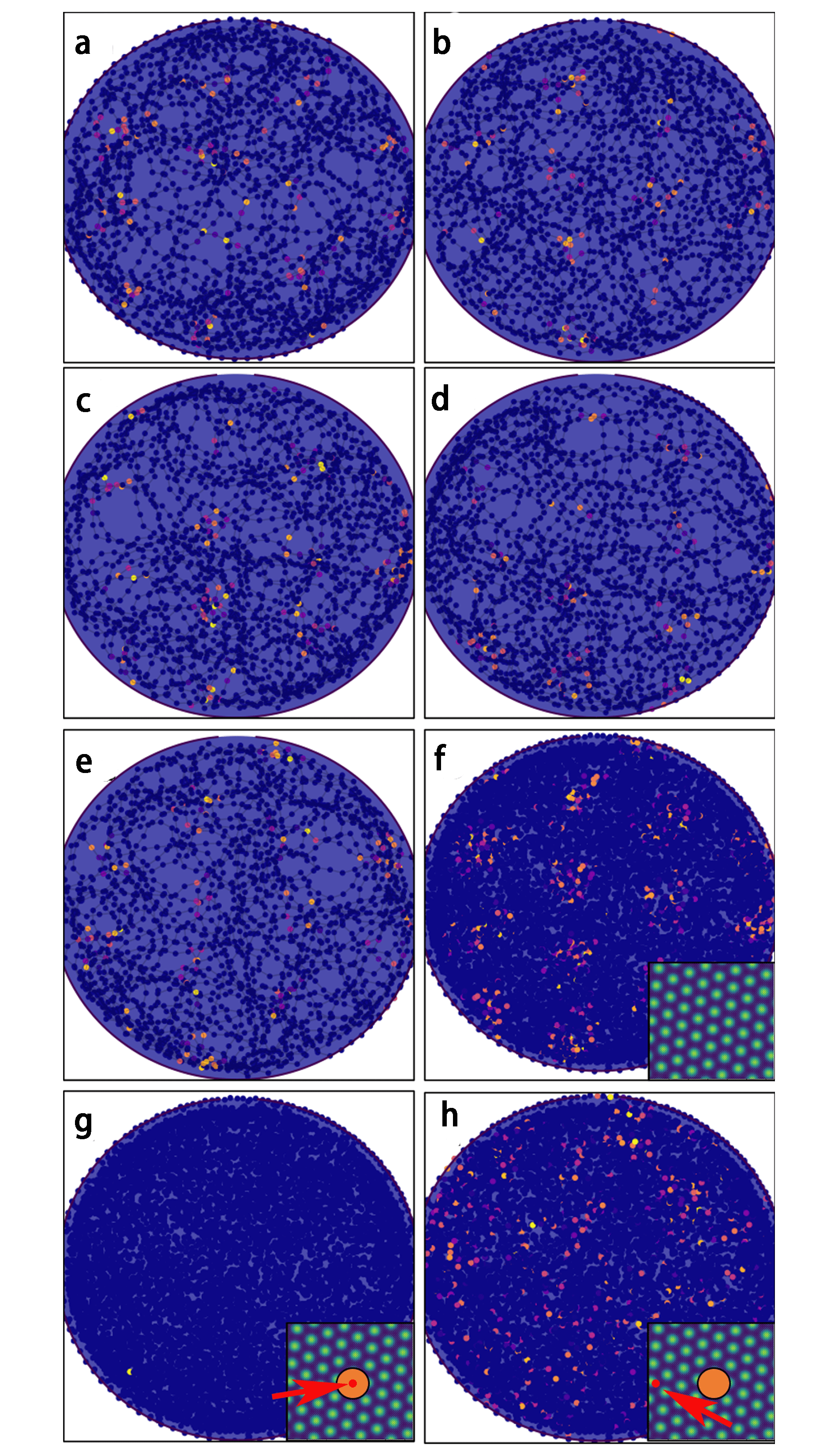}
\caption{\label{fig5} {\bf Single Path Integration Map and Average Path Integration Maps of Healthy/Damaged Neuron Sheets.} (a)$\sim$(e), Single path integration map of healthy neuron sheets for five different trajectories. (f) Average path integration map of the above five shows a clear triangular grid pattern. Inset: grid-like firing pattern in neuron space, $40\times 40$ healthy neuron sheets. (g) The firing of a dead neuron (neuron \#820) is muted in path integration map. Inset: grid-like firing pattern in neuron space, $40\times 40$ damaged neuron sheets (orange damage region $R$ = 7 neurons,$\alpha$=0), red arrow points to the tracking neuron's location (within the damaged region). (h) Firing of a healthy neuron (neuron \#800) doesn't generate a grid -like average path integration map with damage. Inset: same as (g), but the tracking neuron is outside the damaged region.}
\end{figure}
\begin{figure*}
\includegraphics[scale=0.22]{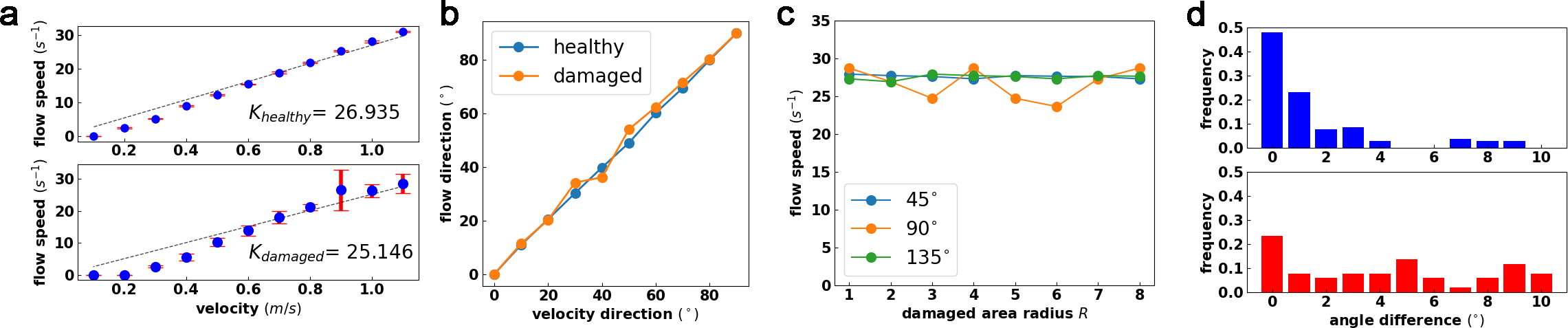}
\caption{\label{fig4} {\bf Relationship of Firing Peak Flow Speed and Velocity Input.} (a) Linear relationship of average flow speed and magnitude of velocity input for undamaged cells. Top is in the healthy neuron sheets and bottom is in the damaged neuron sheets ($R=7,\alpha=0$), the scaling ratios $K$ are given in the figure, and the error bars are the variance of flow speed data, the dashed lines are fitted line with zero intercept. (b) Stability of linear relationship under different velocity inputs directions. The velocity magnitude is $0.7m/s$, and directions are changed from $0^\circ$ to $90^\circ$. The blue curve is the average flow speed direction in the healthy neuron sheets and the orange one is in the damaged neuron sheets ($R=7,\alpha=0$). The direction of average flow speed remains consistent with the direction of velocity input direction.(c) Stability of linear relationship under different damage sizes. The velocity input is $1m/s$ in three directions $(45^\circ, 90^\circ, 135^\circ)$, and the central damage size increases from $R$ = 1 to $R$ = 8 neurons, with the central neurons dead ($\alpha=0$). The overlapping of horizontal lines indicates the average flow speeds are the same if the velocity input magnitudes are the same, regardless of the change of damage size or velocity direction. (d) Angular difference of flow direction and velocity input direction in both healthy and damaged neuron sheets. Top, healthy neuron sheet, and bottom is in the damaged ones. We   recorded the flow direction of 100 firing peaks in both healthy and damaged neuron sheets ($R=7,\alpha=0$), and subtract them by velocity input direction $60^\circ$. The vertical axis is histogram frequency.}
\end{figure*}

In Fig.~\ref{fig3}b, we allowed for a nonzero but weakened connection between neurons in the damaged region and to neurons on the periphery of the undamaged region. A new coefficient $\alpha$ is applied to describe the damage: $W_{damage} = \alpha W_{health}  (0 \leq  \alpha \leq 1 )$. For those neurons that lie in the circular damaged region (assume neuron $i$ in the damaged area $\mathcal{D}$), their presynapses $W_{ji}$ and postsynapses $W_{ij}$ are not the same anymore, and we assume postsynapses (from damaged neurons to healthy neurons) shall be smaller than presynapses (from healthy neurons to damaged neurons).
\begin{equation}
\left\{
             \begin{array}{lr}
              W_{ij} = W_{ji} = W_{0}  \quad (i,j \notin \mathcal{D})  &  \\
              W_{ij} = \alpha W_{ji} = \alpha W_{0} \quad (i \in \mathcal{D}, j \notin \mathcal{D})\\
              W_{ij} = W_{ji} = \alpha W_{0} \quad (i,j \in \mathcal{D})   &  
             \end{array}
\right.
\label{eq6}
\end{equation}

When two neurons are both healthy, their connection weight $W_{ij}$ and $W_{ji}$ ought to be the same as given in Eq.~\ref{eq6}; when one neuron is damaged, we assume it can still accept the signal from other neurons with no reduction but that the signal sent from it will be weaker, this is a more realistic assumption of progressive tau tangle damage than simply killing the neuron.  We achieved this by multiplying the weights by $\alpha$ for all neuronal outputs emerging from within the damaged region. Note that $\alpha$=0 corresponds to dead neurons. 

\subsection{\label{sec:level2}Path Integration}
\subsubsection{\label{sec:level3}Random Walk Generation}
A random walk within a circular enclosure is used to simulate the animal's trajectory in real experiments~\cite{hafting2005microstructure}. Here we use the random walk model of Ref.~\cite{samsonovich2005simple}.
\begin{equation}
\left\{
             \begin{array}{lr}
             \vec{v}_{i+1}=\mu \vec{v}_{i}+\vec{a}_{i}\Delta t &  \\
             \vec{r}_{i+1}=\vec{r}_{i}+\vec{v}_{i} \Delta t \\
             \vec{a}\sim \mathcal N(0,\sigma_a^2)  &  
             \end{array}
\right.
\label{eq7}
\end{equation}

For the $ith$ step, we have velocity  $\vec{v}_{i}$, position  $\vec{r}_{i}$, and acceleration $\vec{a}_{i}$. The acceleration is generated using a Gaussian distribution of random variables, with average of zero and standard deviation $\sigma_a^2=0.5$. The mixing coefficient $\mu$ ($=0.875$ in Table~\ref{tab1}) determines the amount of the current velocity preserved in the next velocity step; this assures a realistically smooth trajectory for which the additional random acceleration boost offers smaller course corrections. We use a small time step $\Delta t =0.1s$ to make sure the change of the model rat’s trajectory is smooth. The velocity is reflected at the boundary, i.e., the component parallel to the boundary is unchanged and the component perpendicular is reversed whenever the model animal would reach the boundary on the next step.  
In most situations, the boundary is reached in-between steps ($|\vec{r}_{i}| < |\vec{R}_{boundary}| < |\vec{r}_{i+1}|$), and thus we recalculate the new position $\vec{r}_{i+1}$ to make it in the reflected direction and have the length $|\vec{r}_{i} \to \vec{R}_{boundary} \to \vec{r}_{i+1}|$ equal $\vec{v}_{i} \Delta t$.

\subsubsection{Path Integration Map}
Now we generate $N$ steps in the random walk path, and each step contains its velocity $\vec{v_i}$ and position $\vec{r_i}$. For each step, the time step $\Delta t =0.1s$ means updating the neuron sheets' signal 200 times ($0.1s/0.5ms =200$), and the whole process uses $\vec{v_i}$  as the velocity input to initiative flow in the firing pattern. We track a single neuron for either damaged or undamaged regions and record their firing rates as the model animal moves to the position $\vec r_i$. The position and single tracking neuron firing rate are plotted together showing the single path integration map (Fig.~\ref{fig5}a). This is exactly the same idea of planting electrode measuring activity of a single neuron in a rat’s dMEC and tracking the firing signal with the rat’s trajectory.\cite{hafting2005microstructure}

\section{Results and Discussion}
\subsection{Linear Grid Pattern Flow Velocity Relationship in Healthy/Damaged Models}

The dynamics of the firing patterns are associated with the velocity input $\vec v$. This velocity changes its direction and magnitude when an animal runs ~\cite{Stangl2018}, which helps us get the actual path information into our future path integration calculation. With zero velocity input, all the firing peaks will be on vertices of a stationary hexagonal lattice. With nonzero velocity input, the model input with direction sensitive cells tiling the grid cell layer initiates a flow of the firing signal, in the opposite direction to velocity input $\vec v$. The linear relationship between the flow speed and velocity input $\vec{\textbf{v}}$ is very important to generate the later accurate path integration (Fig.~\ref{fig5}a$\sim$e). Thus, we introduce the scaling ratio $K$ to describe the relationship:
\begin{equation}
\mbox{flowing speed} =K\cdot\left | \vec{\textbf{v}} \right |  
\label{eq8}
\end{equation}

For damaged neuron sheets in Fig.~\ref{fig3}, we observed the whole driving pattern continues, bypassing the damage. The healthy neurons will still fire normally, while the damaged neurons will be fully muted or weaker in excitatory response. The firing pattern of healthy neurons is not strongly influenced by damaged ones, and still shows partial stability with flow. We quantitatively compare the relationship between in healthy neuron sheets and the damaged one in the following.

Fig.~\ref{fig4}a shows both healthy and damaged neurons sheets retain the linear relationship between average flow speed magnitude and velocity input $\vec{\textbf{v}}$. Clearly, a damaged neuron cluster with say $R = 7$ is pretty big in our $40\times 40$ neuron sheets, but the proportional relationship between average flow speed and physical speed $\vec{\textbf{v}}$ is still close to that of the undamaged neuron sheet. The scaling ratio for the healthy one is $K_{healthy}=26.93 \: m^{-1}$ and for damaged one is  $K_{healthy}=25.146 \: m^{-1}$. 

In Fig.~\ref{fig4}b, we show the average flow directions under different velocity input directions for both healthy and damaged neuron sheets, and clearly the average flow speed direction remains consistency with the velocity input direction. But the damaged one show bigger variance at angle $30^\circ\sim 60^\circ$ than healthy one. In Fig.~\ref{fig4}c, we increase the central damage region from $R = 0$ to $R = 8$ neurons, and compare the flow speed under the velocity inputs of the same magnitude but different directions $(45^\circ, 90^\circ, 135^\circ)$, The overlap of lines indicates that the change of velocity direction doesn't dramatically influence the firing pattern flow, with the same velocity input magnitude ($1m/s$ in Fig.~\ref{fig4}c). And the increased damaged size along the horizontal axis doesn't change the average flow speed a lot.

Those averaged data show good proportionality between flow rate and velocity in both healthy and damaged neuron sheets, regardless of damage size or velocity inputs (magnitudes and directions). However, when we looked into the detailed behavior, the stability and structural coherence of flow for damaged sheets are poor compared those of healthy sheets. 
In Fig.~\ref{fig4}a, the variance (red error bars) of flow speed magnitude in the damaged situation is much bigger than that of healthy ones, meaning the damaged neuron sheets' flow is unstable with fluctuation of flow speed. Then we take record of flow direction of 100 firing peaks (peaks are shown in Fig.~\ref{fig2}d), and subtract them from the velocity input direction ($60^\circ$). Those 100 firing peaks are divided into 10 groups for different velocity magnitude ($0.1m/s\sim 1.0m/s$) in both healthy and damaged neuron sheets. The good linear proportionality shows in Fig.~\ref{fig4}b means the average angle difference should be close to zero. And we noticed that in the healthy neuron sheets, the stability of the triangular lattice is strong and the angle difference is mostly less than $2^\circ$, while in the damaged neuron sheets, the angle difference can up to $10^\circ$. 

In the damaged neuron sheets, the average flow speed proportionality to velocity, but the big variance of direction (and the flow speed fluctuation) in damaged neuron sheets accumulates errors in long-time path integration, which explains why we see the flow pass the damaged hole but the path integration map for cells outside the damaged region cannot generate a triangular lattice (Fig.~\ref{fig5}h). What is interesting is that we found even in healthy neuron sheet, that there are still a few firing peaks with big angle differences (Fig.~\ref{fig4}d top). Those big angle changes in $7^\circ - 9^\circ$ are all from firing peaks under very slow velocity input ($0.1~m/s$), and it indicates too small velocity input is unable to keep firing rate flow's stability either.

\begin{figure*}
\includegraphics[scale=0.15]{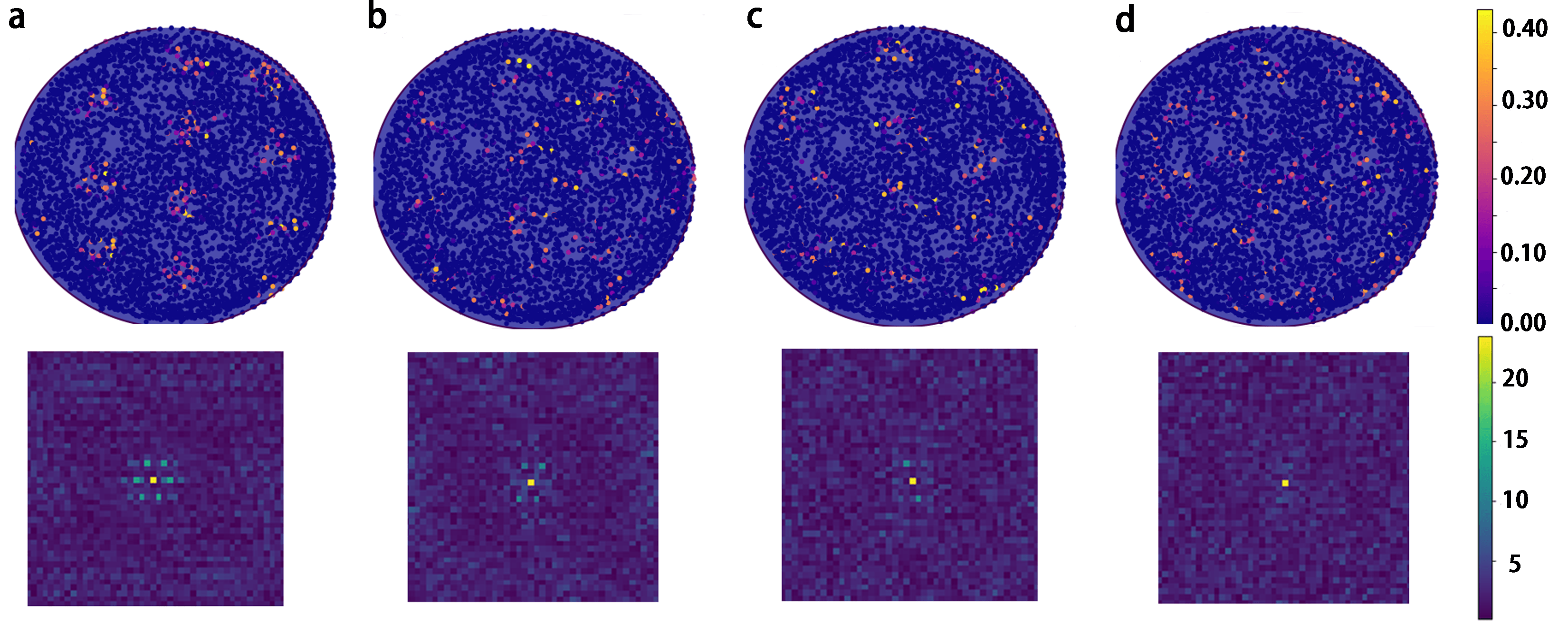}
\caption{\label{fig6} {\bf Average Path Integration Map with Model Damage and Discrete Fourier Transform (DFT).} (a) For neuron \#800, damage coefficient $\alpha$ = 1, which is a healthy neuron sheets, the associated average path integration map shows clear triangular grids. DFT diagram has a hexagonal structures of 6 peaks around the center. (b) Neuron \#800, damage coefficient $\alpha$ = 0.5, damage radius $R$ = 4 neurons. DFT diagram has 4 peaks around the center. (c) Neuron \#820, damage coefficient $\alpha$ = 0.3, damage radius $R$= 2 neurons. DFT diagram has 2 peaks around the center. (d) Neuron \#800, damage coefficient $\alpha$ = 0.4, damage radius $R$= 4 neurons. DFT diagram has 0 peaks around the center. Average path integration maps in (b),(c),(d) are regraded as none-grids path integration map.}
\end{figure*}

\begin{figure*}
\includegraphics[scale=0.15]{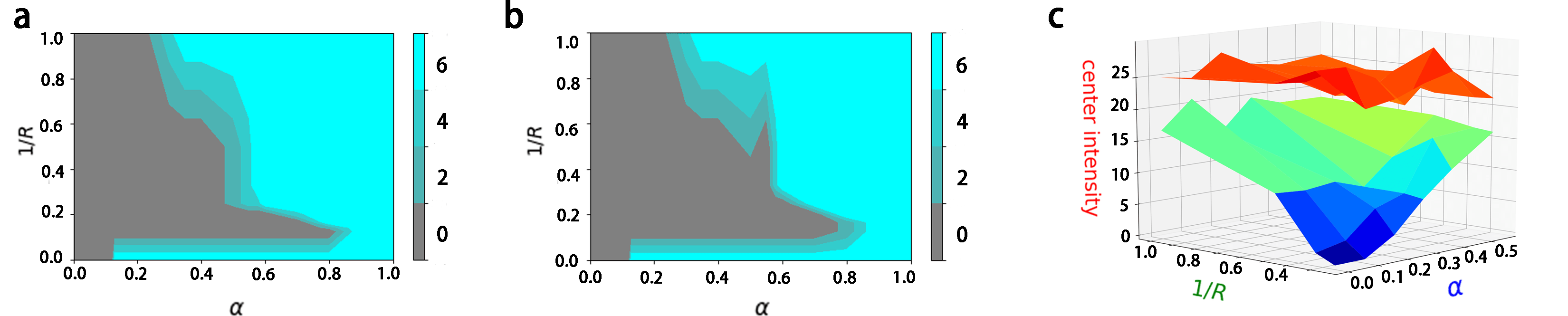}
\caption{\label{fig7} {\bf Phase Diagrams and Fourier Transform Central Peak Intensity.} (a) For neuron \#800 phase diagrams of grid cell order in the $1/R$-$\alpha$ plane. In the teal region we find hexagonal lattice grids in the average path integration map; in the charcoal region there are no grids in the average path integration map region. The two other shaded regions between teal and charcoal correspond to a striped a grid (2 peaks in the DFT), an orthorhombic grid (4 peaks in the DFT).  For $1/R$ = 0, all neurons are damaged $\alpha$, and for $1/R$ = infinity or $\alpha$ = 1 all neurons are healthy. (b) Firing phase diagrams for neuron \#820 phase diagrams in the $1/R$- $\alpha$ plane. The structure is nearly the same as the phase diagram in (a). (c) DFT central peak intensity as a function of $1/R$ and $\alpha$ for undamaged and damaged cells in the region without coherent grid cell firing (charcoal area in (a),(b)). The top surface is for the healthy neuron \#800 (healthy neurons), and the bottom surface is for the \#820 (damaged neuron). Damaged neurons have weak firing signal and bigger variance with changing of $1/R$ and $\alpha$.}
\end{figure*}

\subsection{Path Integration Map for the Healthy Neuron Model}

From the single path integration simulation, we found the firing pattern shows weak grids, more like discrete highlight dots separated in a grid pattern (Fig.~\ref{fig5}a$\sim$e). To achieve a more accurate and clear result, we use the average of five different path integration, for each of them, the same tracking neuron signal starts with the same firing pattern at the very beginning, but follows different trajectories. We add the five sets of firing rate together to generate the path integration map. In Fig.~\ref{fig5}f, we can see the average path integration map present better grids pattern compared with single path integration map. The same idea can be applied if we apply a much longer path, like five times longer, however, increasing the length of path will accumulate integration error because the triangular grids still shows fluctuations. And increasing the length costs longer time for a single simulation. The influence of increasing path length is mentioned in Supporting Information S6.   

When we replaced the healthy neuron sheet with the damaged neuron sheet model, even though damage does not destroy the signal's average flow stability, it influences the path integration in terms of degrading the triangular lattice firing pattern in the random walk on the two dimensional area. In Fig.~\ref{fig5}g,h, we found the damage from killing neurons brought us the worst influence: even a very small region of dead cells can totally destroy the grid from generating. Tracking the dead neuron in the damaged region (Fig.~\ref{fig5}g) shows no firing signal all along the path, the healthy neuron outsider the damaged region are strongly influenced by the damage and cannot generate triangular grids like Fig.~\ref{fig5}f. It seems that the bigger firing pattern fluctuations in damaged neuron sheet accumulate errors in long path. The stability of linear relationship is achieved from single velocity input and not too long simulation ($\sim$ 500ms) while path integration with 1000 steps is about $\sim100s$. 

\subsection{Fourier Transform Analysis and Damage Phase Diagram}

We have found that a small damage region and big damage coefficient $\alpha$ can lead to stable grids, even though too big a damage region size (radius over 7 neurons) or too small a damage coefficient $\alpha$ ($\alpha < 0.2$) still prevent the grids from generating.  Fig.~\ref{fig1}c is a simplified model of 2D $40 \times 40$ neuron sheet with central damage $R$ = 7 neurons, and tracking neuron \#800 is outside the damage region while \#820 is within the region. To get a stable input to the Fourier analysis, we crop the center square region of the path integration map to eliminate boundary effects, and do the discrete Fourier Transformation (DFT) on the truncated position space. For those trajectories with clear enough hexagonal grid structure, the DFT diagrams show six Bragg peaks around the center (Fig.~\ref{fig6}a), where we use the DFT intensity threshold of 15 to quantify the visibility of the peaks to the eye shown in Fig.~\ref{fig6} bottom row,  and those without grids show only one central peak corresponding to the average firing (Fig.~\ref{fig6}d). Between these extremes, we find regions with striped firing (2 non-zero Bragg peaks) and orthorhombic firing (4 non zero Bragg peaks anisotropic in the plane) (Fig.~\ref{fig6}b,c). The application of Bragg peak analysis from the DFT makes it convenient to summarize grid translational and orientational coherence. We want to use the number of Bragg peaks to quantify different levels of grid pattern loses under different damages.  

By studying the non-zero Bragg peaks as a function of $1/R$ and $\alpha$, we can generate the phase diagram shown in Fig.~\ref{fig7}a,b. The grey area is the no grids region and the blue area is grids region. When $\alpha$ is small, or the damage radius is big (corresponding to grey area of Fig.~\ref{fig7}a,b), the loss of a grid like pattern is the worst and there is no Bragg peaks. The opposite is for bigger damage coefficient $\alpha$ or small damage radius $R$, the grid like pattern remains. What we noticed is that there is a borderline to demarcate the grids and no-grids regions. It means that the grid cells show a tolerance of defects so that they can still work well with damages of certain levels. We find with weak synaptic defects ($\alpha > 0.8$), different sizes of damaged region cannot stop the grid like pattern from generating. And with a poor synaptic connection ($\alpha < 0.5$), if the damage region is not too big ($1/R > 0.5, R < 2$), we can still observe a grid pattern (Fig.~\ref{fig7}a). 

The phase diagrams (Fig.~\ref{fig7}a,b) of both neurons are similar in detail, but the discrete Fourier Transformation (DFT) central peak intensities varies. Fig.~\ref{fig7}c is the contour diagram of DFT central peak intensity for neuron within/outside the damage region. For neuron outside the damage region (\#800), the peak intensity is higher and more or less constant in amplitude with increased damage size and strength, while for a neuron in the damaged region (\#820) the central peak intensity diminishes.This peak measures the average firing over the space of the 2-D enclosure. The grid cells within the damaged region are not totally muted and also can generate a grid like pattern (in Fig.~\ref{fig7}b), but its intensity is usually weaker than the grid cells outside the damaged region, and the intensity increases with the distance between the neuron and the damage center. 

The phase diagrams of Fig. 7 display a re-entrant feature at large damage radius $R$.  The reason this exists is clear from passing to the infinite radius limit.  In that case, all neurons have equivalent reduction of their output by the reduction of $\alpha$, so it is guaranteed that the hexagonal peaks will remain until the peak strength in Fourier space is reduced below its critical value. Hence, the critical $\alpha$ value is determined by the inverse of the peak value of $\tilde{W}(\vec{q})$ For the values of $\beta$ and $\gamma$ we use, the analytical estimate for this is, per the Supplemental Information, $\tilde{W}_{max}=4.37$, which gives $\alpha_c = 0.23$ analytically.  The numerical value from \ref{fig7} is somewhat lower ($\alpha_c=0.14$), but given that the Fourier transform estimate in the Supplemental Information does not include the direction dependent offset necessary for generating flow we are comfortable the the argument captures the origins of the re-entrant phase.  

As noted in the introduction, while we anticipate that damage will disrupt the coding of position in the grid cell/place cell network, in this paper we limit our attention to the modification of grid cell symmetry and coherence as a means of providing a map to early detection of neurodegenerative damage. 

\section{Conclusion}
We start from Burak \& Fiete's attractor model of grid cells~\cite{burak2009accurate} to study their firing pattern on much smaller periodic neuron sheets. Smaller neuron sheets can provide good statistics on the firing pattern flow and provide reasonable results for finite and relatively short simulation times. The linearly proportional relationship between the firing pattern flow and animal's velocity input helps build a stable hexagonal lattice in the path integration map. 

Applying simulated damage to the model grid cells, we observe that the firing pattern flow continues but shows bigger fluctuations (in flow speed and direction) in a longer time range. Those fluctuations accumulate errors in path integration and we observe the loss of coherent grid firing.

To identity the tolerance of grid cells to different levels of synaptic damages, we study the Bragg peaks in the Fourier Transformed pattern of position space firing fields for model grid cells, and we have shown that damage to a model grid cell layer parameterized by reduced synaptic output strength $\alpha$ and damage radius $R$ leads to a predictable sequence of reduced grid cell firing symmetry from hexagonal lattice, to orthorhombic lattice, to stripes, and on to no coherent pattern (single central peak).  We find that the central Bragg peak in the region with no coherent grid structure is largely unchanged for grid cells outside the damaged region, but strongly reduced for grid cells in the damaged region. For large area damage, there is a re-entrant transition to the fully hexagonal grid structure. Grid cells can show a tolerance of certain damages. With the help of borderline in phase diagram to identify grids/no-grids region, we can control different levels of synaptic damages on grid cells and study its influence on place coding/decoding.  

The modifications of the orientation of firing patterns  associated with the less ordered structures should be visible in fMRI experiments which can pick up the full six fold symmetric firing pattern in the dMEC for undamaged subjects.  This makes for an important tool in assessing potentially the level of synaptic damage associated with neurodegenerative diseases such as Alzheimer's, that may allow for early diagnosis and the use of small molecule aggregation inhibitor treatments such as anle138b\cite{Antonschmidt2018,Antonschmidt2019}.  

\begin{acknowledgments}
We acknowledge useful conversations with M. Zaki Jawaid at the start of the project and with Rishidev Chaudhuri about the origin of flow of the firing peaks with motion.
\end{acknowledgments}

\nocite{*}

\bibliography{main}

\end{document}